\documentclass[aps,pra, twocolumn, noshowpacs, floatfix]{revtex4}
\usepackage{}

\usepackage{amsfonts}
\usepackage{amssymb}
\usepackage{graphicx}
\usepackage{amsmath}
\usepackage[english]{babel}
\usepackage{color}
\usepackage{epstopdf}

\usepackage{soul}

\begin{document}

\title{Dissipative dynamics in a tunable Rabi dimer with periodic harmonic driving}

\author{Zhongkai Huang$^{1,2\dagger}$, Fulu Zheng$^{3,1\dagger}$, Yuyu Zhang$^{4}$, Yadong Wei$^{3}$, and Yang Zhao$^{1}$\footnote{Electronic address:~\url{YZhao@ntu.edu.sg}.\\ $\dagger$Zhongkai Huang and Fulu Zheng contributed equally to this work.}}
\affiliation{
$^1$Division of Materials Science, Nanyang Technological University, Singapore 639798, Singapore\\
$^2$College of Materials Science and Engineering, Yangtze Normal University, Chongqing 408100, China\\
$^3$School of Physics and Energy, Shenzhen University, Shenzhen 518060, China\\
$^4$Department of Physics, Chongqing University, Chongqing 404100, China}

\begin{abstract}

Recent progress on qubit manipulation allows application of periodic driving signals on qubits. In this study, a harmonic driving field is added to a Rabi dimer to engineer photon and qubit dynamics in a circuit quantum electrodynamics device. To model environmental effects, qubits in the Rabi dimer are coupled to a phonon bath with a sub-Ohmic spectral density. A non-perturbative treatment, the Dirac-Frenkel time-dependent variational principle together with the multiple Davydov D$_2$ {\it Ansatz} is employed to explore the dynamical behavior of the tunable Rabi dimer. In the absence of the phonon bath, the amplitude damping of the photon number oscillation is greatly suppressed by the driving field, and photons can be created thanks to resonances between the periodic driving field and the photon frequency. In the presence of the phonon bath, one still can change the photon numbers in two resonators, and indirectly alter the photon imbalance in the Rabi dimer by directly varying the driving signal in one qubit. It is shown that qubit states can be manipulated directly by the harmonic driving. The environment is found to strengthen the interqubit asymmetry induced by the external driving, opening up a new venue to engineer the qubit states.
\end{abstract}

\maketitle

\section{introduction}

Originally proposed to study the effect of a weak, rapidly rotating magnetic field on an atom  possessing a nuclear spin \cite{Rabi1936, Rabi1937}, the Rabi model represents the simplest interaction between a two-level atom and a light field, and continues to inspire exciting developments in both mathematics and physics \cite{braak_2016}. Since the two-level system in the Rabi model and its variants can describe qubits, the Rabi model has considerable impact on practical applications in quantum computation and information science \cite{alderete_2016}. Recently, the Rabi model has been used to describe various quantum systems, such as microwave and optical-cavity quantum electrodynamics (QED) \cite{raimond_2001}, ion traps \cite{leibfried_2003}, quantum dots \cite{scarlino_2015}, and superconducting qubits in circuit QED \cite{chiorescu_2004}.

Such systems are interesting both for fundamental study of quantum phenomena, such as Landau-Zener transitions \cite{zener_1932, landau_1932}, on the mesoscopic scale, as well as for promising design of future electronic devices. Multiphoton transitions have been proposed and realized in Josephson-junction qubits \cite{schevchenko_2012}. For example, Temchenko {\it et al.}$~$studied the transition between two flux qubits that are biased by independent constant magnetic fluxes, and coupled to each other as well as to an unavoidable dissipative environment \cite{temchenko_2011}. Zheng {\it et al.}$~$reported new results on dynamical photon localization and delocalization in a Rabi dimer model with a dissipative bath \cite{fulu_2018}.

In particular, advances in QED devices and quantum dots make them promising candidates for the exploration of a tunable Rabi model due to their potential scalability and tunable parameters over a broad range \cite{saito_2006, oliver_2005, niemcyzk_2010, scarlino_2015, nalbach_2015}. One way is to apply an external driving field to the cavity \cite{bishop_2010, henriet_2014}, and the other is to impose the driving force on the qubit \cite{wallraff_2004, oliver_2005, schevchenko_2012}. It is more common to tune the energy spacing of the qubit by changing the magnetic flux in the superconducting quantum interference device qubits \cite{Johansson_2009}, or by applying external magnetic field on a spin qubit in a Si/SiGe quantum dot \cite{scarlino_2015}.

Dynamics of the tunable qubits is inevitably influenced by their environments. Environmentally induced fluctuations in the qubit energies are such an example. The qubit-enviroment coupling has been demonstrated to exist in various experiments, such as in a superconducting charge qubit coupled to an on-chip microwave resonator in the strong coupling regime \cite{wallraff_2004}, in a circuit QED device with seven qubits \cite{houck_2008}, and in a circuit QED implementation with a time-dependent transverse magnetic field \cite{viehmann_2013}. QED devices typically work at extremely low temperatures. Dominant noise sources could be modeled by sub-Ohmic spectral densities or 1/$f$ low-frequency noises \cite{xiong_2015}. For example, Egger {\it et al.} have shown that a sub-Ohmic type spectral density can characterize the qubit-bath coupling in a multimode circuit QED setup with hybrid metamaterial transmission lines \cite{egger_2013}. However, effects of qubit-phonon coupling on qubit dynamics have not been sufficiently investigated. Recently, the multiple Davydov D$_2$ {\it Ansatz} has been developed to accurately treat dynamics of the generalized Holstein model with simultaneous diagonal and off-diagonal system-bath coupling \cite{zhou2015polaron, huang_2017}. Influences of qubit-phonon coupling have also been probed in the dissipative Landau-Zener model using our variational approach \cite{huang_lz_2018}.

In a previous study \cite{fulu_2018}, photon delocalization in a Rabi dimer has been studied by employing the multiple Davydov trial states. The external control of the qubit population can be realized in the same experimental setup to engineer photon delocalization in the Rabi dimer by applying two independent magnetic fields on the two qubits. It is thus necessary to extend the successful method to study the dynamics of the tunable Rabi dimer in a dissipative bath \cite{wallraff_2004, nalbach_2017, Braak2011, Zhong2017}. With respect to the dynamics of the tunable Rabi dimer, qubit polarization and photon dynamics affected by qubit-photon and qubit-phonon interactions have not received adequate attention. Here, we continue our endeavor with an accurate treatment of the many-body quantum dynamics in a tunable Rabi dimer.

In this work, we apply periodic harmonic driving to a qubit of the Rabi dimer and explore its effects on the qubit polarizations. We investigate the impacts of qubit-photon coupling and qubit-phonon coupling on the photon dynamics in the tunable Rabi dimer using the multi-$\rm D_2$ {\it Ansatz} with the Dirac-Frenkel variational principle. Good convergence has been obtained using the employed method, justifying the validity of our method.

The remainder of the paper is structured as follows. In Sec.~\ref{methodology}, we present the Hamiltonian and our trial wave function, the multi-$\rm D_2$ {\it Ansatz}. In Sec.~\ref{Numerical results and discussions}, using the Dirac-Frenkel time-dependent variational principle, we proceed to study quantum dynamics of the composite system when one qubit of the Rabi dimer is under harmonic driving and the energy splitting of the other qubit is kept constant. Environmental effects are examined by coupling the qubits to a sub-Ohmic phonon bath. Photon dynamics in the left and right resonators is discussed in Sec.~\ref{photon_dynamics}, and qubit dynamics is investigated in Sec.~\ref{qubit_dynamics}. Conclusions are drawn in Sec.~\ref{Conclusions}.

\section{methodology}
\label{methodology}
\subsection{Hamiltonian of the hybrid system}

\begin{figure}
  \centering
  \includegraphics[scale=1.0]{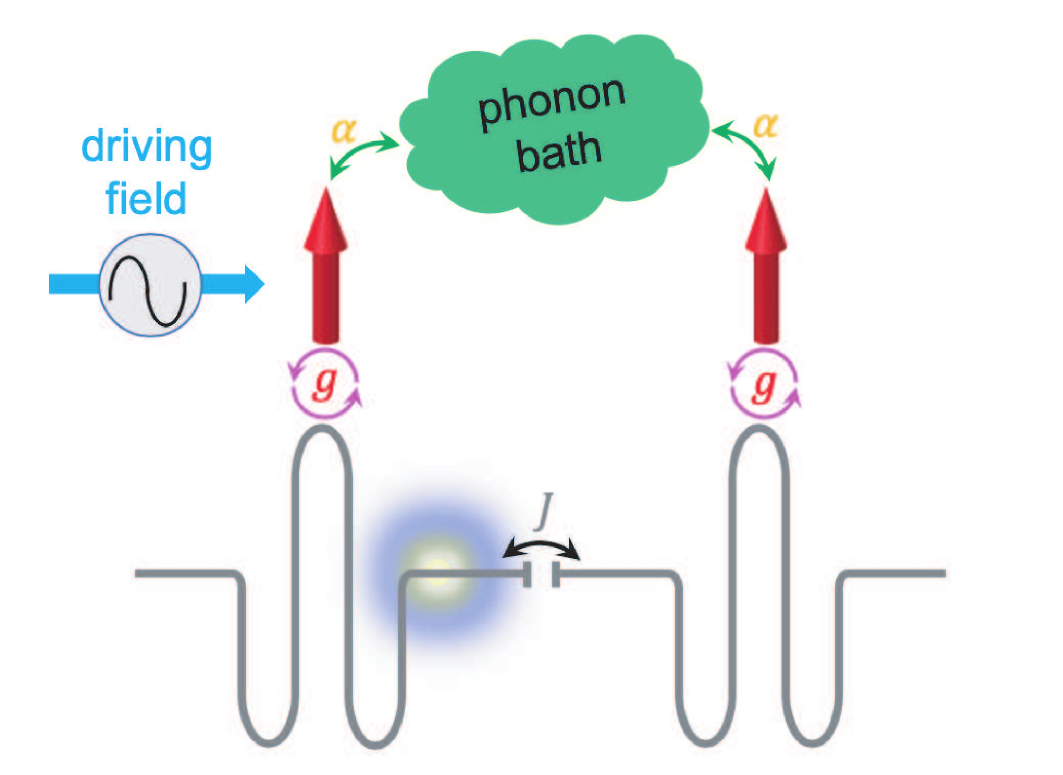}
  \caption{Sketch of a tunable circuit QED system studied in this work. Photons hop between two transmission line resonators with a tunneling rate of $J$. An external periodic driving field is applied to the left qubit. Left (right) qubit is coupled to the photon mode in the left (right) resonator with a coupling strength of $g$. Two qubits interact with a sub-Ohmic phonon bath with a strength of $\alpha$.}\label{Fig1_schem}
\end{figure}

As illustrated in Ref.~\cite{fulu_2018}, we model a Rabi dimer composed of two coupled transmission line resonators with each interacting with a qubit. The environmental effects on the device result from the coupling of the qubits to multimode micromechanical resonators, which are described by a collection of harmonic oscillators. By controlling the magnetic flux through the qubits in the dimer, harmonic driving can be applied onto the qubits, as shown in Fig.~\ref{Fig1_schem}. The Hamiltonian for the hybrid system can be written as
\begin{equation}\label{eq:Htot}
	H=H_{\textrm{RD}}+H_\textrm{B}+H_{\textrm{BQ}}.
\end{equation}

The Rabi dimer can be described by the following Hamiltonian ($\hbar=1$)
\begin{equation}\label{eq:HRD}
	H_{\textrm{RD}}=H^{\textrm{Rabi}}_{\textrm{L}}+H^{\textrm{Rabi}}_{\textrm{R}}-J(a_{\textrm{L}}^{\dagger}a_{\textrm{R}}+a_{\textrm{R}}^{\dagger}a_{\textrm{L}}),
\end{equation}
where $J$ is the photon tunneling amplitude. The energy spacings of left and right qubits are constants in Ref.~\cite{fulu_2018}, and here we make them tunable. To be specific, an external driving field can be independently imposed on each of the two qubits. The left (L) and right (R) Rabi Hamiltonians $H_{i=\textrm{L}/\textrm{R}}^\textrm{Rabi}$ are given by~\cite{Rabi1936, Rabi1937, Braak2011, Zhong2017}

\begin{equation}\label{Hrabi}
	H_{i=\textrm{L}/\textrm{R}}^\textrm{Rabi} = \frac{A_{i}}{2}\cos(\Omega_{i}t+\Phi_{i}) \sigma_{z}^{i} + \omega_{i} a_{i}^{\dagger} a_{i} - g_{i} ( a_{i}^{\dagger} + a_{i} ) \sigma_{x}^{i},\nonumber
\end{equation}
where $\frac{A_{i}}{2}\cos(\Omega_{i}t+\Phi_{i})$ serves as the periodic harmonic driving field on the $i$th qubit, and imply tunable energy spacing of the qubit. $\Phi_L=\Phi_R=0$ is set to simplify the simulations. $\omega_{i}$ is the frequency of the photon mode in the $i$th Rabi system. $\sigma_{x}^{i}$ and $\sigma_{z}^{i}$ are the usual Pauli matrices, and $a_{i}$ ($a_{i}^{\dagger}$) is the annihilation (creation) operator of the $i$th photon mode. $g_{i}$ characterizes the coupling strength between the qubits and the photons. We assume the frequencies of the photon modes and the qubit-photon coupling strengths are identical in the tunable dimer, i.e., $\omega_{\textrm{L}}=\omega_{\textrm{R}}=\omega_{0}$, and $g_{\textrm{L}}=g_{\textrm{R}}=g$.

To study the dynamics of the tunable Rabi dimer in the presence of an environment, we model a phonon bath of $N$ quantum harmonic oscillators by the Hamiltonian $H_\textrm{B}$ and the qubit-phonon coupling by the Hamiltonian $H_{\textrm{BQ}}$,
\begin{equation}\label{Hb}
	H_\textrm{B}=\sum_{k=1}^{N} \omega_{k} b_{k}^{\dagger} b_{k}
\end{equation}
and
\begin{equation}\label{Hbq}
	H_{\textrm{BQ}}=\sum_{k=1}^{N} \phi_{k} (b_{k}^{\dagger}+b_{k})(\sigma_{z}^{\textrm{L}}+\sigma_{z}^{\textrm{R}})
\end{equation}
where $b_{k}$ ($b_{k}^{\dagger}$) is the annihilation (creation) operator of the $k$th bath mode with frequency $\omega_{k}$, and $\phi_{k}$ is the coupling strength between the $k$th mode and the qubits. The qubit-bath coupling is characterized by {a spectral density} function,
\begin{equation}
	J(\omega )=\sum_{k} \phi _{k}^{2} \delta( \omega - \omega _{k} )=2 \alpha \omega _{c}^{1-s} \omega ^{s} e^{-\omega/\omega_{c}},
\end{equation}
with $\omega _{c}$ being the cut-off frequency and the dimensionless parameter $\alpha$ quantifying the qubit-bath coupling strength \cite{WangLu2016}. Since the focus of this work is the application of an external field on the Rabi dimer, we use a weak photon tunneling strength $J$, producing energy levels with small gaps in the spectrum of the Rabi dimer. Low frequency bath modes are crucially important to the dynamics, as these modes may be at resonance with some transitions in the Rabi dimer. Therefore, a sub-Ohmic bath spectral density with $s=0.5$ is chosen in this work.
The logarithmic discretization procedure is adopted to parameterize the low frequency bath modes with balanced numerical accuracy and efficiency~\cite{WangLu2016}. The cut-off frequency for the bath modes is set to $\omega_{c}=\omega_{0}$, and the maximum frequency used in the discretization is $\omega_{\textrm{max}}=20~\omega_{c}$.

If the energies corresponding to frequencies of the photon and phonon modes are high in comparison with the thermal energy $k_BT$, the oscillators are thermally inactive, and thus the dynamics influenced by the bath modes is temperature independent in a wide temperature range \cite{wallraff_2004, chiorescu_2004}. QED devices typically work at extremely low temperatures. Therefore, $T=0$ is adopted in the current study. It is straightforward to include the temperature effects in this methodology by applying Monte Carlo importance sampling \cite{Wang2017}. Simulations at finite temperatures require relatively more computational resources, and will be performed in future studies.

\subsection{The Multi-D$_2$ {\it Ansatz}}

The multiple Davydov D$_2$ {\it Ansatz} with multiplicity $M$ are essentially $M$ copies of the single Davydov D$_2$ {\it Ansatz} \cite{zh_12,zh_97}. It is also known as the multi-${\rm D}_2$ {\it Ansatz}, and has been employed to study static and dynamic properties of various systems, producing excellent numerical efficiency and accuracy in a broad parameter regime in the presence of both diagonal and off-diagonal coupling~\cite{Zhou2016, Chen2017, Wang2017, zhou2015polaron, huang_2017, huang_2017_off, huang_SF_2017, huang_lz_2018, fulu_2018, huang_2018_ac}. In this work, the multi-${\rm D}_2$ {\it Ansatz} is employed to treat accurately both the off-diagonal qubit-photon coupling and the diagonal qubit-phonon system-bath coupling in Eq.~(\ref{eq:Htot}), and can be constructed as
\begin{eqnarray}\label{eq:MD2}
	|{\rm D}_{2}^{M}(t)\rangle &=& \sum_{n=1}^{M} \Big[ A_{n} (t)|\uparrow\uparrow\rangle + B_{n} (t) |\uparrow\downarrow\rangle + C_{n} (t) |\downarrow\uparrow\rangle \nonumber\\
	&&~~~~+ D_{n} (t) |\downarrow\downarrow\rangle \Big] \bigotimes |\mu_{n}\rangle_{\textrm{L}}|\nu_{n}\rangle_{\textrm{R}} |\eta_{n}\rangle_{\textrm{B}},
\end{eqnarray}
where $|\uparrow \downarrow \rangle=| \uparrow \rangle_{\textrm{L}} \otimes | \downarrow \rangle_{\textrm{R}}$ with $\uparrow$ $(\downarrow)$ indicating the up (down) state of the qubits. $|\mu_{n}\rangle_{\textrm{L}}$ and $|\nu_{n}\rangle_{\textrm{R}}$ are coherent states of the photon modes
\begin{eqnarray}
	|\mu_{n}\rangle_{\textrm{L}} & = & \exp\left[\mu_{n} (t) a_{\textrm{L}}^{\dagger}-\mu_{n}^{\ast} (t) a_{\textrm{L}}\right]|0\rangle_{\textrm{L}},\\
	|\nu_{n}\rangle_{\textrm{R}} & = & \exp\left[\nu_{n} (t) a_{\textrm{R}}^{\dagger}-\nu_{n}^{\ast} (t) a_{\textrm{R}}\right]|0\rangle_{\textrm{R}},
\end{eqnarray}
where $|0\rangle_{\textrm{L}(\textrm{R})}$ is the photon vacuum state of the left (right) resonator. $|\eta_{n}\rangle_{\textrm{B}}$ is the coherent state of the phonon bath
\begin{equation}\label{eta}
	|\eta_{n}\rangle_{\textrm{B}} = \exp \left[ \sum_{k}\eta_{nk} (t)  b_{k}^{\dagger}-\eta_{nk}^{\ast} (t) b_{k} \right] |0\rangle_{\textrm{B}}
\end{equation}
with $|0\rangle_{\textrm{B}}$ being the vacuum state of the phonon bath. In Eq.~(\ref{eq:MD2}), $A_{n}(t)$, $B_{n}(t)$, $C_{n}(t)$, $D_{n}(t)$, $\mu_{n}(t)$, $\nu_{n}(t)$, and $\eta_{nk}(t)$ are time-dependent variational parameters to be determined via the time-dependent variational principle. $A_{n}$ is the probability amplitude in the state $|\uparrow\uparrow\rangle|\mu_{n}\rangle_{\textrm{L}}|\nu_{n}\rangle_{\textrm{R}}|\eta_{n}\rangle_{\textrm{B}}$, $\mu_{n}$ ($\nu_{n}$) is the displacement of the left (right) photon mode, and $\eta_{nk}$ is the displacement of the $k$th bath mode.

\subsection{The time-dependent variational principle}

Equations of motion for the variational parameters are derived by adopting the Dirac-Frenkel time-dependent variational principle,
\begin{equation}\label{DiracFrenkel}
 	\frac{d}{dt} \bigg( \frac{\partial L}{ \partial \dot{\alpha}^{*}_{n}} \bigg) - \frac{\partial L}{ \partial \alpha^{*}_{n}} =0.
\end{equation}
where $\alpha_{n}$ are the variational parameters, i.e., $A_{n}(t)$, $B_{n}(t)$, $C_{n}(t)$, $D_{n}(t)$, $\mu_{n}(t)$, $\nu_{n}(t)$, and $\eta_{nk}(t)$. The Lagrangian $L$ is given by
\begin{equation}\label{Lagrangian}
	L = \frac{i}{2} \langle {\rm D}_{2}^{M}(t) | \frac{\overrightarrow{\partial}}{\partial t} - \frac{\overleftarrow{\partial}}{\partial t} | {\rm D}_{2}^{M}(t) \rangle - \langle {\rm D}_{2}^{M}(t) | H | {\rm D}_{2}^{M}(t) \rangle.
\end{equation}
Details of the derivations can be found in Appendix~\ref{Equations of Motion}.

\subsection{Observables}

Employing the Dirac-Frenkel time-dependent variational principle with the multi-D$_2$ {\it Ansatz}, we investigate the bath induced dynamics of a Rabi dimer with specific contributions from individual bath modes presented explicitly. The time evolution of photon numbers in two resonators is given by
\begin{eqnarray}
	N_{\textrm{L}}(t) & = & \langle{\rm D}_{2}^{M}(t)| a_{\textrm{L}}^{\dagger} a_{\textrm{L}} |{\rm D}_{2}^{M}(t) \rangle \nonumber\\
	& = & \sum_{l,n}^{M} \Big[ A_{l}^{\ast}(t) A_{n}(t)  + B_{l}^{\ast}(t) B_{n}(t)  + C_{l}^{\ast}(t) C_{n}(t) \nonumber\\
	&&~~~~~~~ + D_{l}^{\ast}(t) D_{n}(t) \Big] \mu_{l}^{\ast}(t) \mu_{n}(t)  S_{ln}(t) , \\
	N_{\textrm{R}}(t) & = & \langle{\rm D}_{2}^{M}(t)|a_{\textrm{R}}^{\dagger}a_{\textrm{R}}|{\rm D}_{2}^{M}(t) \rangle \nonumber \\
	& = & \sum_{l,n}^{M} \Big[ A_{l}^{\ast}(t) A_{n}(t)  + B_{l}^{\ast}(t) B_{n}(t)  + C_{l}^{\ast}(t) C_{n}(t) \nonumber\\
	&&~~~~~~~ + D_{l}^{\ast}(t) D_{n}(t) \Big] \nu_{l}^{\ast}(t) \nu_{n}(t)  S_{ln}(t) ,
\end{eqnarray}
where $S_{ln} (t)$ is the Debye-Waller factor
\begin{eqnarray}
	S_{ln} & = &  \exp\left[\mu_{l}^{\ast}(t) \mu_{n}(t)-\frac{1}{2} |\mu_{l}(t)|^2-\frac{1}{2}|\mu_{n}(t)|^2\right] \cdot \nonumber\\
	&& \exp\left[ \nu_{l}^{\ast}(t) \nu_{n}(t) - \frac{1}{2} |\nu_{l}(t)|^2 - \frac{1}{2} |\nu_{n}(t)|^2 \right] \cdot \nonumber\\
	&&\exp \left[ \sum_{k} \Big( \eta_{lk}^{\ast}(t) \eta_{nk}(t) - \frac{1}{2} |\eta_{lk}(t)|^2 - \frac{1}{2} |\eta_{nk}(t)|^2 \Big) \right]. \nonumber\\
\end{eqnarray}
The time evolution of the photon imbalance is $Z(t)=N_{\textrm{L}}(t)-N_{\textrm{R}}(t)$ and that of the total photon number is ${N}(t)=N_{\textrm{L}}(t)+N_{\textrm{R}}(t)$. These quantities are used to characterize photon localization and delocalization.

In addition to photon dynamics, the time evolution of the qubit states is recorded during the simulations by measuring the time evolution of the qubit polarization via
\begin{eqnarray}
	\langle\sigma_{z}^{\textrm{L}}(t)\rangle & = & \langle{\rm D}_{2}^{M}(t)|\sigma_{z}^{\textrm{L}}|{\rm D}_{2}^{M}(t) \rangle \nonumber \\
	& = & \sum_{l,n}^{M} \Big[ A_{l}^{\ast}(t)A_{n}(t) + B_{l}^{\ast}(t)B_{n}(t) \nonumber\\
	&&- C_{l}^{\ast}(t)C_{n}(t) - D_{l}^{\ast}(t)D_{n}(t) \Big] S_{ln}(t),\\
	\langle\sigma_{z}^{\textrm{R}}(t)\rangle & = & \langle{\rm D}_{2}^{M}(t)|\sigma_{z}^{\textrm{R}}|{\rm D}_{2}^{M}(t) \rangle \nonumber \\
	& = & \sum_{l,n}^{M} \Big[ A_{l}^{\ast}(t)A_{n}(t) - B_{l}^{\ast}(t)B_{n}(t) \nonumber\\
	&&+ C_{l}^{\ast}(t)C_{n}(t) - D_{l}^{\ast}(t)D_{n}(t) \Big] S_{ln}(t).
\end{eqnarray}
As given in Hamiltonian (\ref{eq:Htot}), the qubits serve as a bridge to connect the photon and the phonon modes, transferring bath-induced impacts to the photons. Combining influences from the photons and the bath, our calculated qubit dynamics reflects the complex interactions between the photon modes and the phonon bath.

Thanks to the methodology adopted here, the temporal evolution of the phonon bath can also be obtained explicitly. To reveal the participation of individual phonon modes in the Rabi dimer dynamics, we calculate the population on the $k$th mode as follows
\begin{eqnarray}
	N_{k}^{\textrm{B}}(t) &=& \langle{\rm D}_{2}^{M}(t)| b_{k}^{\dagger} b_{k}|{\rm D}_{2}^{M}(t) \rangle \nonumber \\
	&=& \sum_{l,n}^{M} \Big[ A_{l}^{\ast}(t) A_{n}(t) + B_{l}^{\ast}(t) B_{n}(t) + C_{l}^{\ast}(t) C_{n}(t) \nonumber\\
	&&~~~~~~~~+ D_{l}^{\ast}(t) D_{n}(t) \Big] \eta_{lk}^{\ast}(t) \eta_{nk}(t) S_{ln}(t).
\end{eqnarray}
Through interacting with the qubits, the phonon bath gradually gains sufficient energy from the Rabi dimer to affect the dynamics of the photons and the qubits. In return, the influences of the QED system on the bath modes can be investigated by calculating the populations dynamics $N_{k}^{\textrm{B}}(t)$.

\section{Results and discussion}
\label{Numerical results and discussions}

Used as comparison is a case where the qubits and the photons are at resonance in the absence of a driving field. The parameter set chosen is $A_L/\omega _{0}=A_R/\omega _{0}=1$, $\Omega_L=\Omega_R=0$, and $\Phi_L=\Phi_R=0$. In the absence of a phonon bath and a driving field, the photon dynamics is determined by the cooperation between $J$ and $g$. A relatively weak photon tunneling rate $J=0.05\omega_0$ is adopted here, and the quadrature-quadrature coupling between the two photon modes is neglected~\cite{Rossatto2016, WangYM2016}. According to recent experiment~\cite{Raftery2014}, ultrastrong qubit-photon coupling (USC) is preferred, thus the qubit-photon coupling strength is set to $g=0.3~\omega_{0}$. The combined effects of $J$ and $g$ used here lead to photon delocalization over two resonators in a bare Rabi dimer~\cite{Hwang2016}. We choose a qubit-bath coupling strength of $\alpha=0.1$ to model the phonon effect. A two-qubit gate has been studied by H\"anggi {\it et al.}, and an ac field was acted upon one of the qubits to induce a time-dependent level splitting \cite{hanggi_2005}. Here, a harmonic driving field is applied to the left qubit without loss of generality. Similar to previous experimental and theoretical work~\cite{Raftery2014, fulu_2018}, a fully localized photon state is prepared by pumping $N(0)=20$ photons into the left resonator while keeping the right one in a photon vacuum with $\mu_1(t=0)=\sqrt{20}$ and $\mu_{n\neq1}(t=0)=\nu_n(t=0)=0$. The qubits in the two resonators start to evolve from their down states with $A_n(t=0)=B_n(t=0)=C_n(t=0)=D_{n\neq1}(t=0)=0$ and $D_1(t=0)=1$. The phonon bath is initially in a vacuum state with $\eta_{nk}(t=0)=0$. Validity of our approach has been extensively tested in the previous work, and calculations are performed with a sufficiently large multiplicity $M$ in this work. The multiplicity $M=8$ and number of phonon modes $N_{\rm bath}=0$ is set for cases of $\alpha=0$, while $M=6$ and $N_{\rm bath}=60$ is used for cases of $\alpha=0.1$. The reader is referred to Ref.~\cite{fulu_2018} for details of the validity study.

As the driving force is exerted on the left qubit, the site energy on the qubit can be directly controlled by following parameters: the driving field strength $A_L$, the harmonic driving frequency $\Omega_L$, and the ratio between them $R=A_L/\Omega_L$. The effect of the driving field will be transferred to the left photon mode via the qubit-photon coupling $g$. In the proposed QED device shown in Fig.~\ref{Fig1_schem}, the photons initially created will hop between two resonators with a tunneling rate of $J$.

\subsection{Photon dynamics}
\label{photon_dynamics}

\begin{figure}
  \centering
  \includegraphics[scale=1.0]{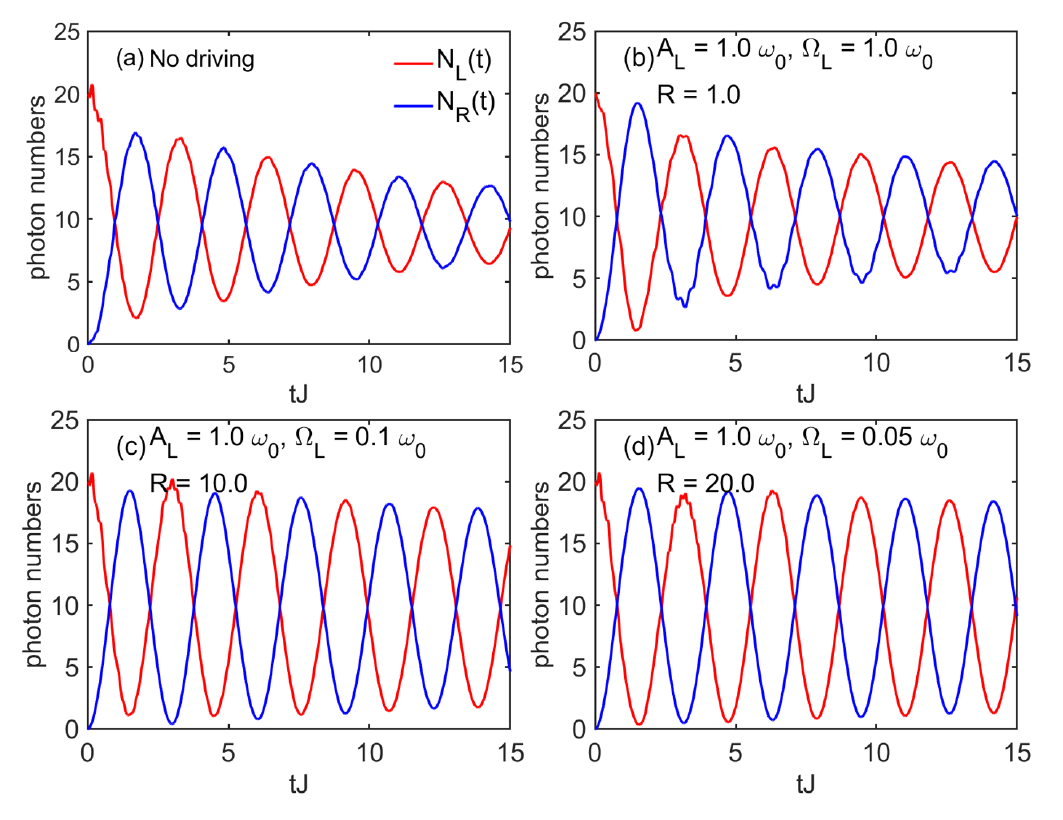}
  \caption{Time evolution of the photon numbers in the left ($N_{\textrm{L}}(t)$) and right ($N_{\textrm{R}}(t)$) resonators without phonon bath ($\alpha=0$). Various fields are applied to the left qubit: (a)$A_L=A_R=\omega_{0}$, $\Omega_L=\Omega_R=0$, (b)$A_L=\omega_{0}$ and $\Omega_L=\omega_{0}$, (c)$A_L=\omega_{0}$ and $\Omega_L=0.1~\omega_{0}$, and (d)$A_L=\omega_{0}$ and $\Omega_L=0.05~\omega_{0}$. There is no field on the right qubit ($A_R=0$) in (b)-(d). The multiplicity used in the corresponding calculations is $M=8$. } \label{Fig2_photon_no_bath}
\end{figure}

We first study the photon dynamics in the composite system. Shown in Fig.~\ref{Fig2_photon_no_bath} are the time-dependent photon numbers in two resonators in the absence of the phonon bath. The photons hop between the resonators due to the inter-resonator tunneling rate $J$. For the case without a driving field and a phonon bath, as shown in Fig.~\ref{Fig2_photon_no_bath}(a), the photon number in an individual resonator approaches half of the initial total photon number, indicating photon delocalization with quasiequilibration in two resonators at long times \cite{fulu_2018}. Due to the qubit-photon coupling, the oscillation amplitude decays as time evolves. The decrease of the oscillation amplitude can be seen as a purity loss in one Rabi model or the decoherence in the Rabi dimer \cite{hanggi_2005}. Upon applying a harmonic driving field to the left qubit, the external field provides energy to the Rabi dimer and the oscillation amplitude damping of the photon numbers is suppressed, as displayed in Figs.~\ref{Fig2_photon_no_bath}(b)-(d). H\"anggi {\it et al.}$~$discovered that the decoherence in the two-qubit system at low temperatures can be significantly slowed due to the application of the harmonic driving, and the purity loss depends on the ratio $R$ between the driving field strength $A_L$ and the driving frequency $\Omega_L$ \cite{hanggi_2005}. As expected, Figs.~\ref{Fig2_photon_no_bath}(b)-(d) show that the decay of the oscillation amplitude of the photon number is decelerated by an increasing ratio $R$ with the fixed driving field strength $A_L=\omega_{0}$.

\begin{figure}
  \centering
  \includegraphics[scale=1.0]{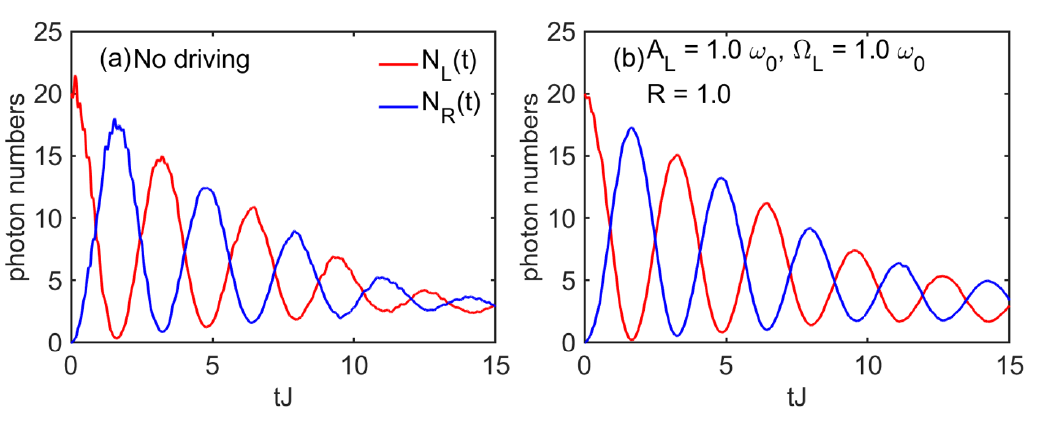}
  \caption{Time evolution of the photon numbers in the left ($N_{\textrm{L}}(t)$) and right ($N_{\textrm{R}}(t)$) resonators with phonon bath ($\alpha=0.1$). Following fields are added to the left qubit: (a) $A_L=A_R=\omega_{0}$, $\Omega_L=\Omega_R=0$, (b) $A_L=\omega_{0}$, $A_R=0$, and $\Omega_L=\omega_{0}$. The multiplicity $M$ and the number of phonon modes $N_{\rm bath}$ used in the related calculations are: $M=6$ and $N_{\rm bath}=60$.} \label{Fig3_photon_bath_R1_10_20_A1}
\end{figure}

\begin{figure}
  \centering
  \includegraphics[scale=1.0]{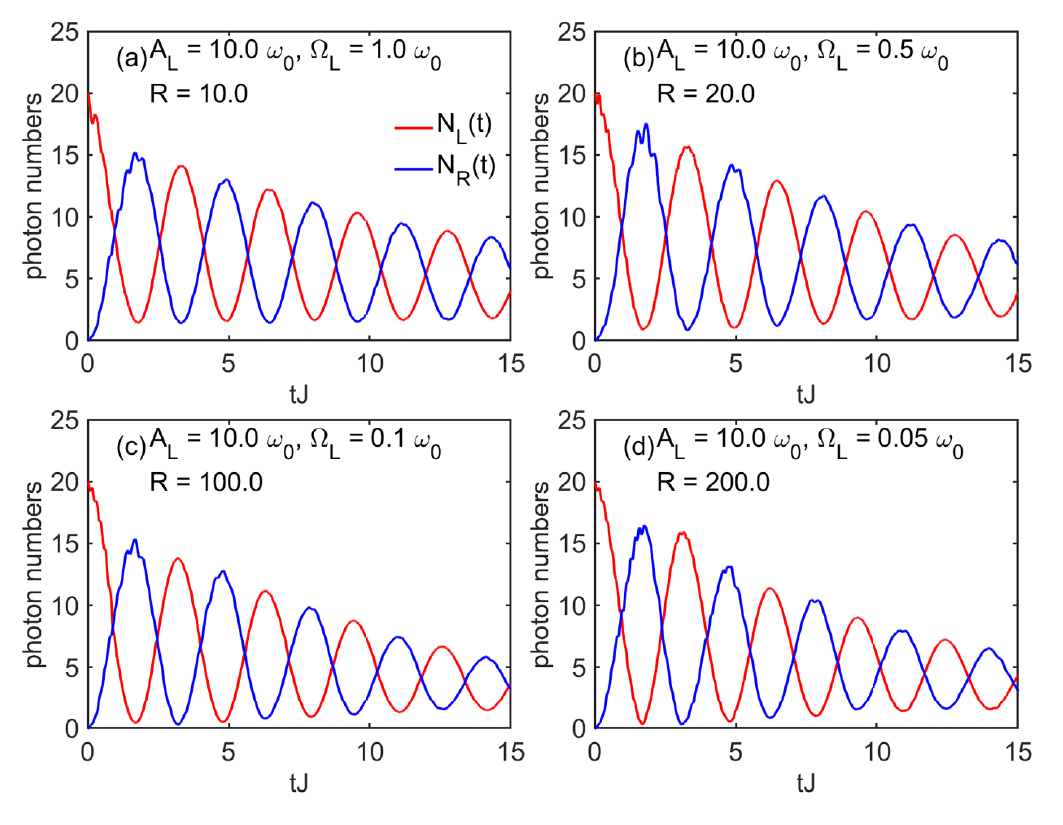}
  \caption{Time evolution of the photon numbers in the left ($N_{\textrm{L}}(t)$) and right ($N_{\textrm{R}}(t)$) resonators with phonon bath ($\alpha=0.1$). Following fields are added to the left qubit: (a)$A_L=10.0~\omega_{0}$, $\Omega_L=\omega_{0}$, (b)$A_L=10.0~\omega_{0}$ and $\Omega_L=0.5~\omega_{0}$, (c)$A_L=10.0~\omega_{0}$ and $\Omega_L=0.1~\omega_{0}$, and (d)$A_L=10.0~\omega_{0}$ and $\Omega_L=0.05~\omega_{0}$. There is no field on the right qubit ($A_R=0$). The multiplicity $M$ and the number of phonon modes $N_{\rm bath}$ used for the figure are: $M=6$ and $N_{\rm bath}=60$.} \label{Fig4_photon_bath_R10_20_100_200_A10}
\end{figure}

Upon turning on the qubit-bath coupling, the photon numbers in two resonators deviate dramatically from those without a phonon bath, as shown in Fig.~\ref{Fig2_photon_no_bath} and Fig.~\ref{Fig3_photon_bath_R1_10_20_A1}. In the absence of the driving field, the oscillation amplitude of the photon numbers in Fig.~\ref{Fig3_photon_bath_R1_10_20_A1}(a) drops faster than those in Fig.~\ref{Fig2_photon_no_bath}(a), since the phonon bath absorbs the energy from the qubit via the qubit-bath coupling, leaving few photons active in the resonators at long times. Our results are in agreement with the conclusion that the environmental noise may lead to strong purity loss in a Rabi model \cite{hanggi_2005}. The field applied in Fig.~\ref{Fig3_photon_bath_R1_10_20_A1}(b) is the same as that for Fig.~\ref{Fig2_photon_no_bath}(b). It is found that a driving field with strength $A_L=\omega_{0}$ is incapable to keep the photons active, and can not significantly offset the photon dynamics.

In order to offset the dissipative effects from the phonon bath, the driving field strength is increased to a larger value of $A_L=10~\omega_0$, and the ratio $R=A_L/\Omega_L$ varies from $10$ to $200$, as shown in Fig.~\ref{Fig4_photon_bath_R10_20_100_200_A10}.
The phononic effects on the photon dynamics under a strong field are much weaker than those under a weak field. As shown in Fig.~\ref{Fig4_photon_bath_R10_20_100_200_A10}, a larger ratio $R$ implies a lower harmonic driving frequency, and the oscillation amplitude of the photon numbers generally decays faster as $R$ increases in the presence of the phonon bath. In contrast, the decay of the oscillation amplitude becomes slower as $R$ grows larger when the phonon bath is absent, as shown in Fig.~\ref{Fig2_photon_no_bath}. This is because the photon bath has only one mode with the frequency of $\omega_0$ while the phonon bath includes more low-frequency modes. The sub-Ohmic phonon bath has strong dissipative effects on the photon dynamics for low-frequency drving. Detailed phononic effects on the photon numbers can be better understood by the corresponding population dynamics of the bath modes, as will be presented in Fig.~\ref{phonon}. It is found that a strong driving field can significantly suppress the decay of oscillation amplitude of the photon numbers even in the presence of the phonon bath.

\begin{figure}
  \centering
  \includegraphics[scale=1.0]{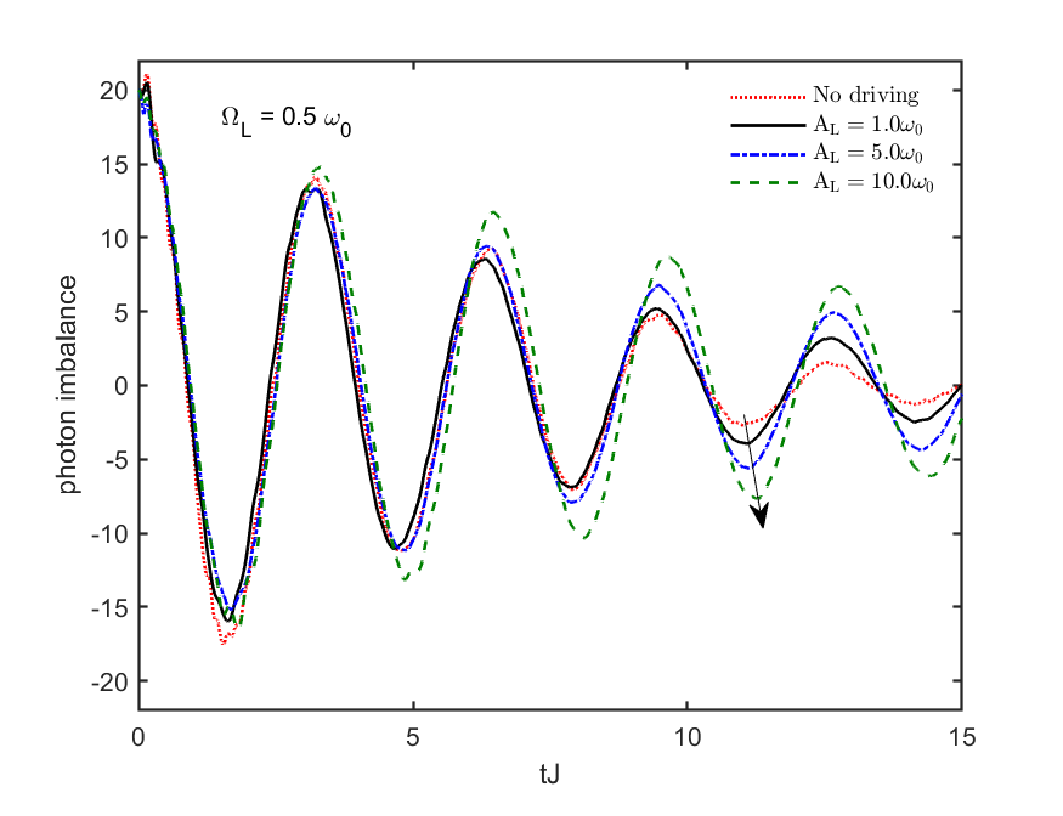}
  \caption{Time evolution of the photon imbalance $Z(t)$ with phonon bath ($\alpha=0.1$). Following fields are added to the left qubit: $A_L=A_R=\omega_{0}$ and $\Omega_L=\Omega_R=0$ for the red dotted line, $A_L=\omega_{0}$ for the black solid line, $A_L=5.0~\omega_{0}$ for the blue dash-dotted line, and $A_L=10.0~\omega_{0}$ for the green dashed line. The driving field frequency is $\Omega_L=0.5~\omega_{0}$. There is no field on the right qubit ($A_R=0$) for nonzero $A_L$. The arrow indicates changes of oscillation periods due to increasing driving field amplitudes. The multiplicity $M$ and the number of phonon modes $N_{\rm bath}$ used in the corresponding simulations are: $M=6$ and $N_{\rm bath}=60$.} \label{Fig5_photon_imbalance}
\end{figure}

Next, we study the time evolution of the photon imbalance in the presence of the qubit-bath coupling of $\alpha=0.1$. As shown in Fig.~\ref{Fig5_photon_imbalance}, the oscillation amplitude for the no driving case decreases to 2 around $tJ=14$. In comparison, slower decay is presented as the driving field strength increases with a harmonic driving frequency of $\Omega_L=0.5~\omega_{0}$. This is because the oscillation amplitude of the photon number evolution decreases more gradually when the ratio of field-strength and frequency is raised. Varying the driving field strength is similar to tuning the qubit-photon coupling strength and the tunneling rate, leading to a more delocalized photon state \cite{hau_2011}.
As extensively studied by Savel'ev {\it et al.},$~$manipulation of one qubit can be realized by applying an ac signal to an adjacent qubit coupled with it via interqubit interaction \cite{savel_2005, savel_2010, savel_2012_1, savel_2012_2}. We have a more sophisticated system with two photon modes and one phonon bath, but we still can indirectly alter the photon imbalance between the two resonators by changing the driving signal in one qubit coupled to its related resonator. The photon dynamics in one resonator can be further controlled by manipulation of the driving field on the qubit coupled to the neighbouring resonator. With a fixed qubit-phonon coupling strength, it is the competitive qubit-photon interaction $g$ and inter-resonator photon tunneling $J$ that determine the photon dynamics in a Rabi dimer. Hausinger and Grifoni have shown that the external driving field on qubit can amplify the effects of qubit-photon coupling \cite{hau_2011}. Therefore, $J$ appears relatively weaker compared to the effective $g$ as the driving field strength is increased \cite{Hwang2016}, leading to longer period oscillations of the photon imbalance as shown in Fig.~\ref{Fig5_photon_imbalance}. This longer period for stronger $A_{L}$ is detectable at long times as illustrated by the arrow at around $tJ=11.3$.

\begin{figure}
  \centering
  \includegraphics[scale=1.0]{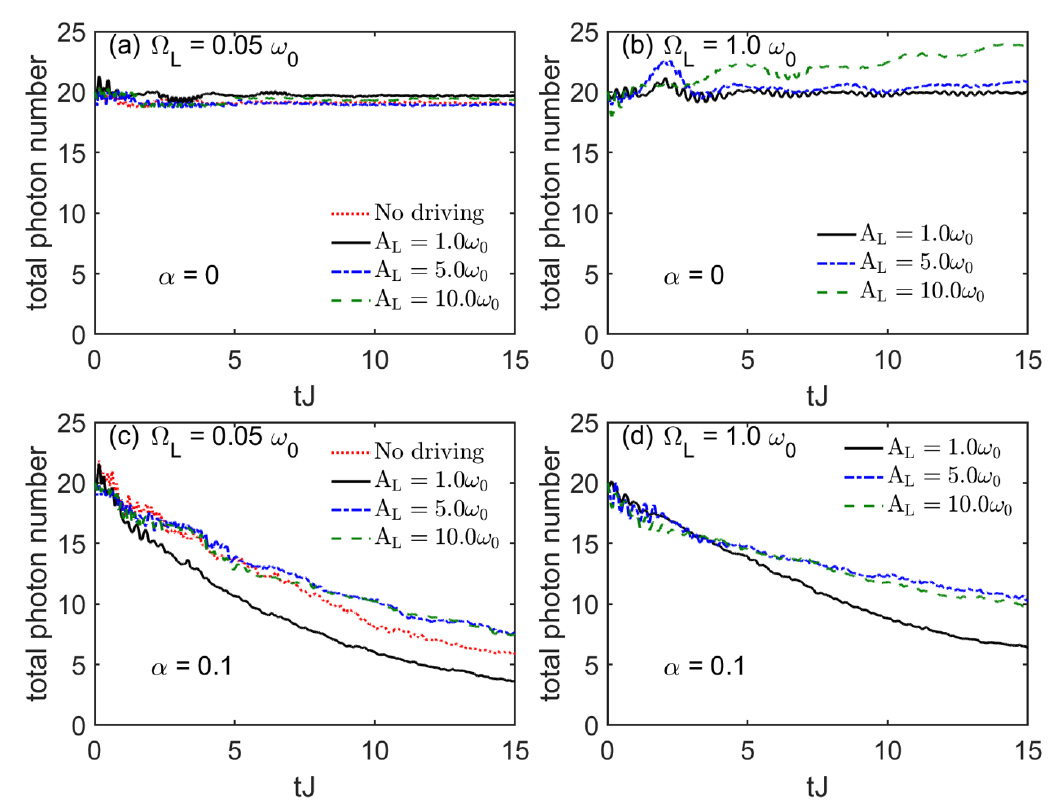}
  \caption{Time evolution of total photon number. The qubit-bath couping $\alpha$ and the driving field frequency $\Omega_L$ are: (a) $\alpha=0$ and $\Omega_L=0.05~\omega_0$, (b) $\alpha=0$ and $\Omega_L=1.0~\omega_0$, (c) $\alpha=0.1$ and $\Omega_L=0.05~\omega_0$, (d) $\alpha=0.1$ and $\Omega_L=1.0~\omega_0$. 
  In each subplot, following fields are added to the left qubit: $A_L=A_R=\omega_{0}$ and $\Omega_L=\Omega_R=0$ for the red dotted line, $A_L=1.0~\omega_{0}$ for the black solid line, $A_L=5.0~\omega_{0}$ for the blue dash-dotted line, and $A_L=10.0~\omega_{0}$ for the green dashed line. There is no field on the right qubit ($A_R=0$) for nonzero $A_L$. The multiplicity $M$ and the number of phonon modes $N_{\rm bath}$ used in the calculations are: $M=8$ and $N_{\rm bath} = 0$ for (a) and (b), and $M=6$ and $N_{\rm bath}=60$ for (c) and (d).} \label{Fig6_tot}
\end{figure}

We then explore the time evolution of the total photon number to study photon creation in the absence of the phonon bath, as shown in Figs.~\ref{Fig6_tot} (a) and \ref{Fig6_tot} (b). Initially, the total photon number is $20$, and all the photons are found in the left resonator. Without the external field, the photon number decreases at short times because the right photon mode is in its initial vacuum state. The high-frequency oscillations in the time evolution of the total photon number have the same frequency as the photons, which is much larger than the photon tunneling amplitude $J=0.05~\omega_0$. At long times, though the qubit-photon coupling $g=0.3~\omega_0$ is strong enough to ensure photon delocalization over two resonators in a bare Rabi dimer, the photon numbers stays almost constant during inter-resonator tunneling, as presented by the red dotted line in Fig.~\ref{Fig6_tot}(a). The application of various harmonic driving fields scarcely affects the total photon number if the driving field frequency $\Omega_L$ is off-resonant with the photon frequency $\omega_0$. With a fixed photon frequency $\omega_0$, we can vary the frequency of the harmonic driving $\Omega_L$. If $\Omega_L$ is in resonance or near resonance with $\omega_0$, the total photon number is found to increase with time, as shown in Fig.~\ref{Fig6_tot}(b). Even in simple driving models with a single qubit, rich physics has sometimes been uncovered thanks to the resonance between the periodic driving field and the energy splitting in the qubit systems \cite{oliver_2005, son_2009, hau_2011, cao_2007}. In a strongly driven superconducting qubit, a fringe pattern can be found in the peaks of interqubit transition probability around multiphoton resonance positions \cite{oliver_2005, son_2009}. Cao {\it et al.}$~$studied a driven spin-boson model with the driving field frequency in resonance with the tunneling in the two level system, and found a damped oscillation for weak driving and an undamped, large-amplitude coherent oscillation for strong driving \cite{cao_2007}. In our model, after the application of the harmonic driving of $\Omega_{L} = \omega_0$ to the left qubit, the total photon number is kept larger than the initial total photon number due to the resonance between the periodic driving field and the photon frequency, as displayed in Fig.~\ref{Fig6_tot}(b). In contrast, the addition of a phonon bath leads to the decrease of total photon number, as shown in Figs.~\ref{Fig6_tot}(c) and (d). Even in the presence of the phonon bath, the total photon number can be compensated by applying a driving field of a sufficiently large amplitude, comparing to that without the field. As indicated in Figs.~\ref{Fig6_tot}(c) and (d), the total photon numbers can be more easily controlled using a resonant driving field than using an off-resonant one.

\subsection{Qubit dynamics}
\label{qubit_dynamics}

\begin{figure}
  \centering
  \includegraphics[scale=1.0]{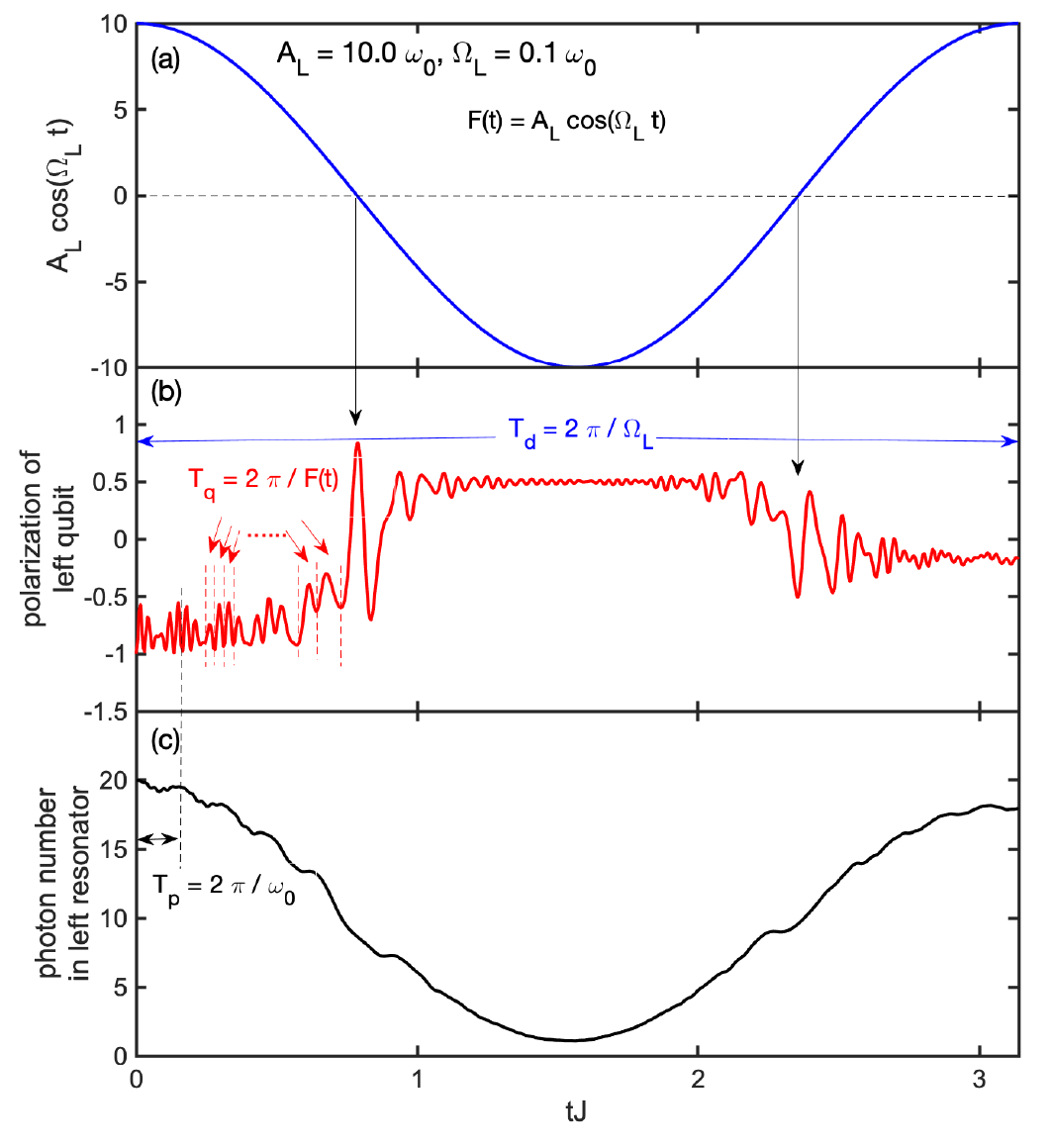}
  \caption{(a) Harmonic driving field, (b) time evolution of the left qubit polarization $\langle\sigma_{z}^{\textrm{L}}(t)\rangle$, and (c) time evolution of the photon numbers in the left ($N_{\textrm{L}}(t)$) resonator. $A_L=10.0~\omega_{0}$ and $\Omega_L=0.1~\omega_{0}$. There is no field on the right qubit ($A_R=0$). There is no phonon bath. The multiplicity used in the related calculations is $M=8$.} \label{qubit_polarization_2}
\end{figure}

In addition to the photon dynamics, the time evolution of the qubit polarization can be monitored. As shown in Fig.~\ref{qubit_polarization_2}, the qubit dynamics can be directly tuned by the external driving field. When the phonon bath is absent, the left qubit is controlled by the driving field, the photon mode, and the photon tunneling. Here we choose a driving field with $A_L=10~\omega_0$ and $\Omega_L=0.1~\omega_0$, and examine the influence of the field and the photon mode on the polarization of the left qubit. The qubit is initially in its down state. As time evolves, the mixing effects lead to segmented oscillations in the left qubit polarization, as displayed in Fig.~\ref{qubit_polarization_2}(b). There are at least four types of energy contributions to the oscillation of the left qubit polarization: the energy of driving field with frequency $\Omega_L$, the photon energy with frequency $\omega_0$, the photon tunneling energy $2J$, and the splitting energy $F(t)=A_L\cos(\Omega_Lt)$ in the left qubit. It can be found that the driving field determines the pattern with the frequency of $\Omega_L$, by comparing Figs.~\ref{qubit_polarization_2}(a) and (b). In the first plateau of Fig.~\ref{qubit_polarization_2}(b), the left qubit polarization oscillates with small amplitudes and the time-dependent frequency of $F(t)$, as indicated by the red arrows. Once the energy difference between the up and down states of the left qubit approaches zero, the left qubit flips from the down (up) state to the up (down) state as denoted by the black down arrows, leading to the second (third) plateau in the overall trend. After flipping at $tJ=\pi/4$, the splitting $F(t)$ becomes larger, and prevents further flipping until $F(t)$ is close to zero again at $tJ=3\pi/4$. In each plateau, the qubit polarization oscillates with a small amplitude and a high frequency due to the combined effects of photon oscillations and the changing qubit splitting energy. This small amplitude can be understood as follows. The photons flow from the left (right) to the right (left) resonator, as shown in Fig.~\ref{qubit_polarization_2}(c). Too few photons in the left resonator are available to affect the qubit polarization at around $tJ=\pi/2$, leading to small amplitude oscillations in the second plateau of the qubit polarization.

\begin{figure}
  \centering
  \includegraphics[scale=1.0]{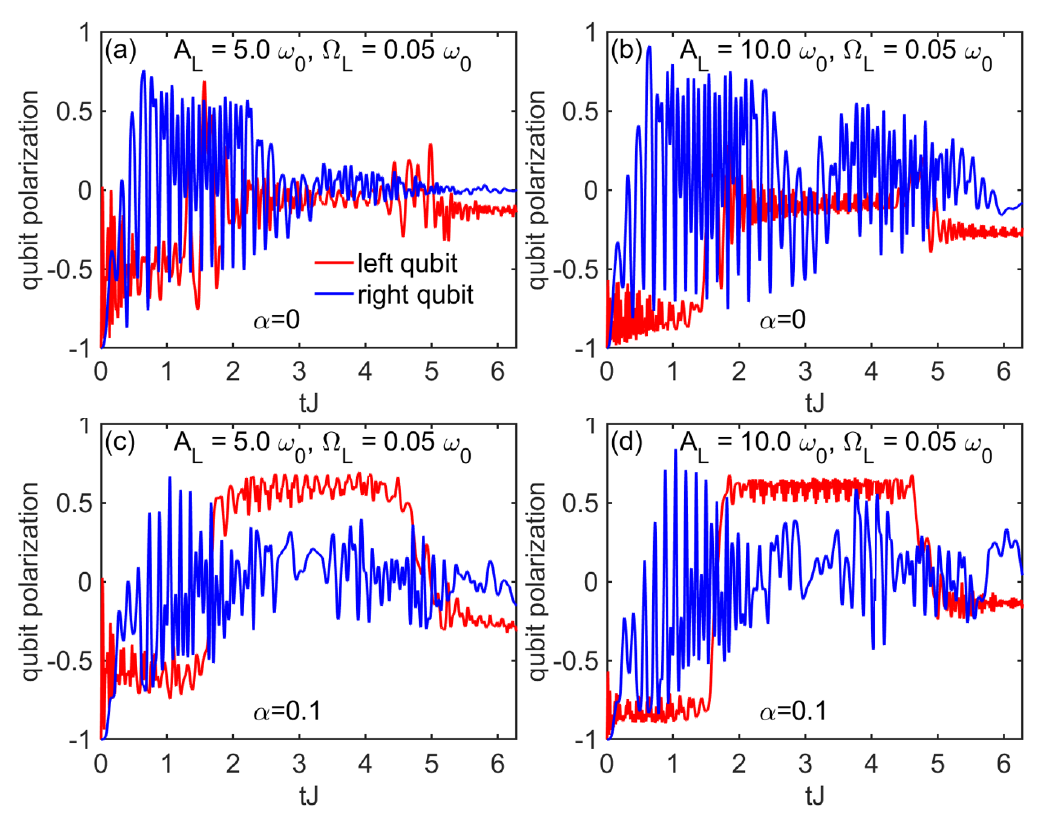}
  \caption{Time evolution of the qubit polarization $\langle\sigma_{z}^{\textrm{L}}(t)\rangle$ and $\langle\sigma_{z}^{\textrm{R}}(t)\rangle$. The qubit-bath coupling $\alpha$ and the driving field parameters are: (a) $\alpha=0$, $A_L=5.0~\omega_{0}$ and $\Omega_L=0.05~\omega_{0}$; (b) $\alpha=0$, $A_L=10.0~\omega_{0}$ and $\Omega_L=0.05~\omega_{0}$; (c) $\alpha=0.1$, $A_L=5.0~\omega_{0}$ and $\Omega_L=0.05~\omega_{0}$; and (d) $\alpha=0.1$, $A_L=10.0~\omega_{0}$ and $\Omega_L=0.05~\omega_{0}$.  
  There is no field on the right qubit ($A_R=0$). The multiplicity $M$ and the number of phonon modes $N_{\rm bath}$ used in the calculations are: $M=8$ and $N_{\rm bath}=0$ for (a) and (b), and $M=6$ and $N_{\rm bath}=60$ for (c) and (d).}
  \label{Fig8_qubit}
\end{figure}

As shown Fig.~\ref{Fig8_qubit}, the external driving creates asymmetry between the left and right qubit polarization. In Figs.~\ref{Fig8_qubit} (a) and \ref{Fig8_qubit}(b), in the absence of a phonon bath, the interqubit asymmetry (i.e., the difference between the blue and red curves) grows as the field strength is increased. The oscillation amplitude of the right qubit polarization (blue) vanishes at shorter times in Fig.~\ref{Fig8_qubit} (a) than that in Fig.~\ref{Fig8_qubit} (b). Segmented oscillations (square-wave like patterns) appear in the left qubit polarization (red) and are strengthened as the field strength is increased. In Figs.~\ref{Fig8_qubit} (c) and \ref{Fig8_qubit} (d), when the bath is present, the segmented oscillation shows a higher oscillation amplitude comparing to its counterpart without a phonon bath. In particular, the qubit polarization shows a time periodicity of $2\pi/\Omega_L$ under a strong driving field. In Figs.~\ref{Fig8_qubit} (c) and \ref{Fig8_qubit} (d), the asymmetry between the left and right qubit polarization becomes more obvious, when the phonon bath is present. The phonon bath is diagonally coupled to the qubits and can be treated as a bias on the qubits. The bath can then trap the qubit in its up or down state, resulting in more pronounced asymmetry between the left and right qubits. It can also be understood as follows. Due to the driving field, it is more difficult for the left qubit to flip than the right qubit, as the left qubit flips only when $F(t) = 0$. The results of qubit polarization indicate that we can manipulate the qubit state directly by harmonic driving. Moreover, the asymmetry between the two qubits can be strengthened by the environmental phonons, leading to a new venue to engineer the qubit states.

\begin{figure}
  \centering
  \includegraphics[scale=1.0]{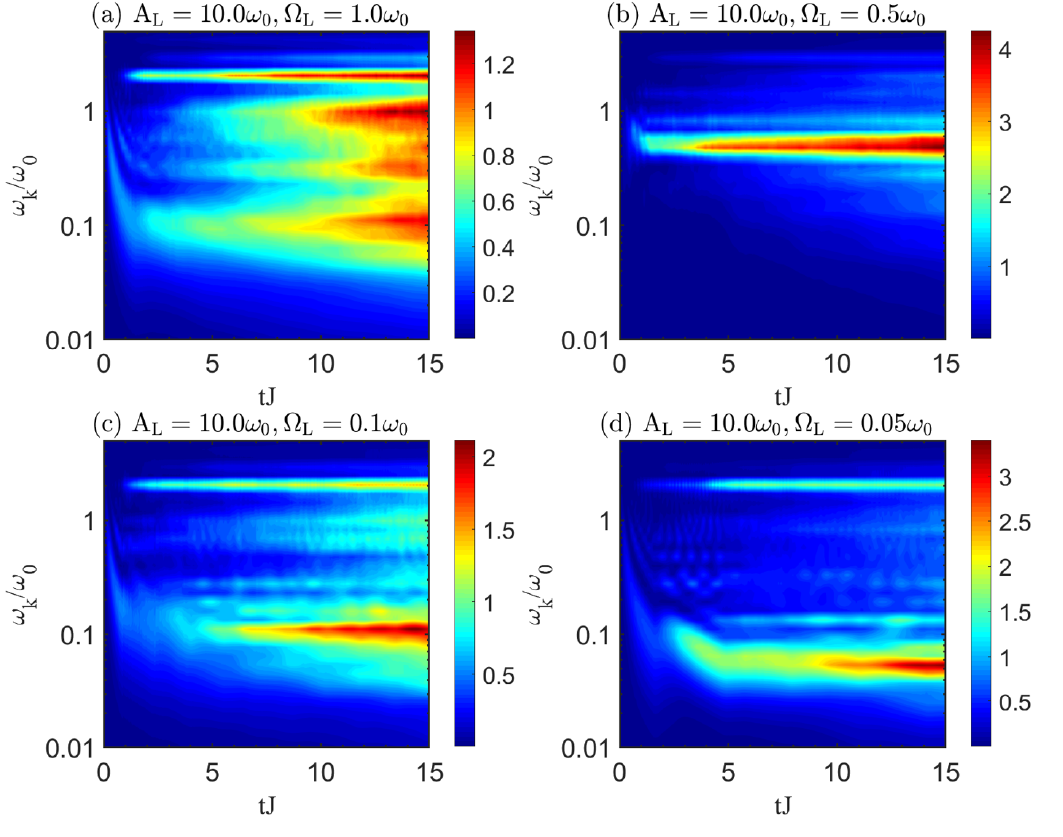}
  \caption{Population dynamics of the bath modes. Following fields are added to the left qubit: (a) $A_L=10.0~\omega_{0}$ and $\Omega_L=\omega_{0}$, (b) $A_L=10.0~\omega_{0}$ and $\Omega_L=0.5~\omega_{0}$, (c) $A_L=10.0~\omega_{0}$ and $\Omega_L=0.1~\omega_{0}$, and (d) $A_L=10.0~\omega_{0}$ and $\Omega_L=0.05~\omega_{0}$. There is no field on the right qubit ($A_R=0$). The multiplicity $M$ and the number of phonon modes $N_{\rm bath}$ used in the calculations are: $M=6$ and $N_{\rm bath}=60$.}
\label{phonon}
\end{figure}

In order to gauge the participation of individual phonon bath modes in the tunable Rabi dimer dynamics, we study population evolution of the bath modes. Thanks to our wave function based method, detailed phonon dynamics is available to shed light on the interplay between the electronic and phononic degrees of freedom (DOFs). In contrast, the phonon DOFs are traced out when constructing the reduced density matrix in the density matrix based methods, and explicit information of the bath dynamics is lost \cite{hanggi_2005, cao_2007}. Without the driving field, two qubits can freely flip and the total photon number decreases continuously, exciting a large number of phonon modes via the the qubit-phonon coupling \cite{fulu_2018}. After the application of the harmonic driving, interesting physics can be obtained from detailed phonon dynamics. As shown in Fig.~\ref{phonon}(a), the phonon bath is at its vacuum state at $t=0$. At short times, there is an energy influx into the modes with frequencies near $\omega_k=1.5~\omega_0$ \cite{fulu_2018}. After $tJ=2$, there is an increase in the population of $\omega_k=2.0~\omega_0$, which is attributed to the energy passed by the two qubits from their coupled photon modes with the frequency of $\omega_0$ to the phonon modes. Since the left qubit flipping is determined by the driving field, the population of phonons that are in resonance with the driving field frequency $\Omega_L=1.0~\omega_0$ grows larger. Low-frequency phonons are gradually excited at long times, because the sub-Ohmic phonon bath is featured at low-frequency regimes. As presented in Figs.~\ref{phonon}(b)-(d), the phonon modes in resonance with periodic driving field frequency $\Omega_L$ dominate the phonon dynamics, if $\Omega_L$ is smaller than $\omega_0$. This is because the qubits are directly coupled to the phonon bath, and the left qubit continuously receives external energy from the strong field $A_L=10\omega_0$ and switches its orientation at a low frequency $\Omega_L<\omega_0$. In Fig.~\ref{phonon}(b), phonon dynamics is dominated by active phonons with frequencies around $0.5~\omega$. Among the the four cases shown in Fig.~\ref{phonon}, more low-frequency phonon modes are found in Figs.~\ref{phonon}(c) and (d) to contribute to the dissipative effect. Therefore, a smaller oscillation amplitude of the photon numbers in the left and the right resonator is observed at $tJ = 15$ in Figs.~\ref{Fig4_photon_bath_R10_20_100_200_A10}(c) and (d), comparing to those in Figs.~\ref{Fig4_photon_bath_R10_20_100_200_A10}(a) and (b).

\section{Conclusion}
\label{Conclusions}
In our previous work \cite{fulu_2018}, we studied the intriguing role played by the qubit-phonon coupling in engineering photon delocalization in a dissipative Rabi dimer in the absence of external fields. In this work, we extend our formulism to investigate the influence of a harmonic driving field in the vicinity of one of the qubits on the dynamics of the composite system. Following the Dirac-Frenkel time-dependent variationa principle, the photon and the qubit dynamics is probed by employing the multi-D$_2$ {\it Ansatz}. The external harmonic driving field can provide energy to the Rabi dimer and slow the amplitude damping of photon number oscillations induced by qubit-photon coupling in the absence of a phonon bath. Especially, the total photon number is found to be larger than, or equal to, its initial value due to the resonance between the periodic driving field and the photon frequency. In the presence of the phonon bath, the photon numbers are dramatically reduced and the reduction can be partially compensated by strong driving fields. It is revealed that the qubit polarization can be tuned by the harmonic driving. Low-frequency segmented oscillations can be found in the time evolution of the left qubit polarization if the driving field frequency is not high. Environmental effects can strengthen the qubit state asymmetry induced by the driving field. Finally, our phonon dynamics analyses based on the multi-D$_2$ {\it Ansatz} can successfully identify the contribution of specific phonon modes to the time evolution of the photon numbers and the qubit polarization. Recently, simultaneous manipulation of the two qubits has been proposed in a system of two superconducting flux qubits interacting with each other through their mutual inductance \cite{xia_2009}. Work on simulating photon and qubit dynamics is in progress by applying separate external fields on the two qubits.

\section*{Acknowledgments}
The authors thank Cao Xiufeng for useful discussion. Support from the Singapore National Research Foundation through the Competitive Research Programme (CRP) under Project No.~NRF-CRP5-2009-04 and from the Singapore Ministry of Education Academic Research Fund Tier 1 (Grant Nos.$~$RG106/15, RG102/17, and RG190/18) is gratefully acknowledged.

\appendix
\section{The time dependent variational approach}
\label{Equations of Motion}

The Dirac-Frenkel variational principle results in equations of motion of the variational parameters as follows,

\begin{eqnarray}
&&i \sum_{n=1}^{M} ( \dot{A}_{n}  +   A_{n} \Xi_{nl} ) S_{ln} \nonumber\\
&=& \sum_{n=1}^{M} \Bigg\{ A_{n} \cdot \Big[ \frac{\Delta_{L}(t) + \Delta_{R}(t)}{2} + \Pi_{nl} + 2\sum_{k}\phi_{k}\left(\eta_{lk}^{*}+\eta_{nk} \right) \Big] \nonumber\\
&& - g \Big[ (\mu_{l}^{\ast}+\mu_{n})C_{n} + (\nu_{l}^{\ast}+\nu_{n})B_{n} \Big] \Bigg\} S_{ln},
\end{eqnarray}

\begin{eqnarray}
&& i \sum_{n=1}^{M} ( \dot{B}_{n} + B_{n} \Xi_{nl} ) S_{ln} \nonumber\\
&=& \sum_{n=1}^{M} \Bigg\{ B_{n} \cdot \Big[ \frac{\Delta_{L}(t) - \Delta_{R}(t)}{2} + \Pi_{nl} \Big] - g \Big[ (\mu_{l}^{\ast}+\mu_{n})D_{n} \nonumber\\
&& + (\nu_{l}^{\ast}+\nu_{n})A_{n} \Big] \Bigg\} S_{ln},
\end{eqnarray}

\begin{eqnarray}
&& i \sum_{n=1}^{M} ( \dot{C}_{n} + C_{n} \Xi_{nl} ) S_{ln}\nonumber\\
&=& \sum_{n=1}^{M} \Bigg\{ C_{n} \cdot \Big[ \frac{-\Delta_{L}(t) + \Delta_{R}(t)}{2} + \Pi_{nl} \Big] - g \Big[ (\mu_{l}^{\ast}+\mu_{n})A_{n} \nonumber\\
&&+ (\nu_{l}^{\ast}+\nu_{n})D_{n} \Big] \Bigg\} S_{ln},
\end{eqnarray}

\begin{eqnarray}
&& i \sum_{n=1}^{M} ( \dot{D}_{n} + D_{n} \Xi_{nl} ) S_{ln} \nonumber\\
&=& \sum_{n=1}^{M} \Bigg\{ D_{n} \cdot \Big[ \frac{-\Delta_{L}(t) - \Delta_{R}(t)}{2} + \Pi_{nl} - 2 \sum_{k} \phi_{k} ( \eta_{lk}^{\ast} + \eta_{nk}) \Big] \nonumber\\
&&-g \Big[ (\mu_{l}^{\ast}+\mu_{n})B_{n} + (\nu_{l}^{\ast}+\nu_{n})C_{n} \Big] \Bigg\} S_{ln},
\end{eqnarray}

\begin{eqnarray}
&& i \sum_{n=1}^{M} \Big[ ( A_{l}^{\ast} \dot{A}_{n} + B_{l}^{\ast} \dot{B}_{n} + C_{l}^{\ast} \dot{C}_{n} + D_{l}^{\ast} \dot{D}_{n} ) \mu_{n} \nonumber\\
&& + \Theta_{nl}^{a} ( \dot\mu_{n} + \mu_{n} \Xi_{nl}) \Big] S_{ln} \nonumber\\
&=& \sum_{n=1}^{M} \Bigg\{ \Big[ \frac{\Delta_{L}(t)}{2} \Theta_{nl}^{b} + \frac{\Delta_{R}(t)}{2} \Theta_{nl}^{c} \Big] \mu_{n} \nonumber \\
&& + \Theta_{nl}^{a} \big( \omega_{L} \mu_{n} - J \nu_{n} + \mu_{n} \Pi_{nl} \big) \nonumber \\
&& - g \Theta_{nl}^{d} \big[ 1 + \mu_{n} (\mu_{l}^{\ast}+\mu_{n}) \big] - g \Theta_{nl}^{e} \mu_{n} (\nu_{l}^{\ast}+\nu_{n})\nonumber \\
&& + 2 (A_{l}^{\ast} A_{n} - D_{l}^{\ast} D_{n}) \mu_{n} \sum_{k} \phi_{k} ( \eta_{lk}^{\ast} + \eta_{nk}) \Bigg\} S_{ln}, \nonumber \\
\end{eqnarray}

\begin{eqnarray}
&& i \sum_{n=1}^{M} \Big[ ( A_{l}^{\ast} \dot{A}_{n} + B_{l}^{\ast} \dot{B}_{n} + C_{l}^{\ast} \dot{C}_{n} + D_{l}^{\ast} \dot{D}_{n} ) \nu_{n} \nonumber\\
&& + \Theta_{nl}^{a} ( \dot\nu_{n} + \nu_{n} \Xi_{nl}) \Big] S_{ln} \nonumber\\
&=& \sum_{n=1}^{M} \Bigg\{ \Big[ \frac{\Delta_{L}(t)}{2} \Theta_{nl}^{b} + \frac{\Delta_{R}(t)}{2} \Theta_{nl}^{c} \Big] \nu_{n} \nonumber \\
&& + \Theta_{nl}^{a} ( \omega_{R} \nu_{n} - J \mu_{n} + \nu_{n} \Pi_{nl} ) \nonumber \\
&& - g \Theta_{nl}^{d} \nu_{n} (\mu_{l}^{\ast}+\mu_{n}) - g \Theta_{nl}^{e} \Big[ 1 + \nu_{n} (\nu_{l}^{\ast}+\nu_{n}) \Big]\nonumber \\
&& + 2 (A_{l}^{\ast} A_{n} - D_{l}^{\ast} D_{n}) \nu_{n} \sum_{k} \phi_{k} ( \eta_{lk}^{\ast} + \eta_{nk}) \Bigg\} S_{ln}, \nonumber \\
\end{eqnarray}

\begin{eqnarray}
&& i \sum_{n=1}^{M} \Big[ ( A_{l}^{\ast} \dot{A}_{n} + B_{l}^{\ast} \dot{B}_{n} + C_{l}^{\ast} \dot{C}_{n} + D_{l}^{\ast} \dot{D}_{n} ) \eta_{nk^{\prime}} \nonumber\\
&& + \Theta_{nl}^{a} ( \dot{\eta}_{nk^{\prime}} + \eta_{nk^{\prime}} \Xi_{nl} ) \Big] S_{ln} \nonumber\\
&=& \sum_{n=1}^{M} \Bigg\{ \Big[ \frac{\Delta_{L}(t)}{2} \Theta_{nl}^{b} + \frac{\Delta_{R}(t)}{2} \Theta_{nl}^{c} \big] \eta_{nk^{\prime}} \nonumber \\
&& + \Theta_{nl}^{a} ( \omega_{k^{\prime}} \eta_{nk^{\prime}} +  \eta_{nk^{\prime}} \Pi_{nl} ) \nonumber \\
&& - g \Theta_{nl}^{d} \eta_{nk^{\prime}} (\mu_{l}^{\ast}+\mu_{n}) - g \Theta_{nl}^{e} \eta_{nk^{\prime}} (\nu_{l}^{\ast}+\nu_{n})\nonumber\\
&+& 2 (A_{l}^{\ast} A_{n} - D_{l}^{\ast} D_{n}) \Big[ \phi_{k^{\prime}} + \eta_{nk^{\prime}} \sum_{k} \phi_{k} ( \eta_{lk}^{\ast} + \eta_{nk}) \Big] \Bigg\} S_{ln}, \nonumber \\
\end{eqnarray}
where the auxiliary terms are
\begin{eqnarray}
\Xi_{nl} &=& \dot\mu_{n}\mu_{l}^{\ast}-\frac{1}{2} \dot\mu_{n} \mu_{n}^{\ast} -\frac{1}{2} \mu_{n} \dot\mu_{n}^{\ast} \nonumber\\
&+& \dot\nu_{n}\nu_{l}^{\ast}-\frac{1}{2} \dot\nu_{n} \nu_{n}^{\ast} -\frac{1}{2} \nu_{n} \dot\nu_{n}^{\ast} \nonumber\\
&+& \sum_{k} (\dot{\eta}_{nk} \eta_{lk}^{\ast}- \frac{1}{2} \dot{\eta}_{nk} \eta_{nk}^{\ast}- \frac{1}{2} \eta_{nk} \dot{\eta}_{nk}^{\ast}),
\end{eqnarray}

\begin{eqnarray}
\Pi_{nl} &=&(\omega_{L}\mu_{l}^{\ast}\mu_{n}+\omega_{R}\nu_{l}^{\ast}\nu_{n}) - J(\mu_{l}^{\ast}\nu_{n}+\nu_{l}^{\ast}\mu_{n}) \nonumber\\
&+& \sum_{k} (\omega_{k} \eta_{lk}^{\ast} \eta_{nk}),
\end{eqnarray}

\begin{eqnarray}
\Delta_{L}(t) = A_{L}\cos(\Omega_{L}t+\Phi_{L}),
\end{eqnarray}

\begin{eqnarray}
\Delta_{R}(t) = A_{R}\cos(\Omega_{R}t+\Phi_{R}),
\end{eqnarray}

\begin{eqnarray}
\Theta_{nl}^{a} = A_{l}^{\ast} A_{n} + B_{l}^{\ast} B_{n} + C_{l}^{\ast} C_{n} + D_{l}^{\ast} D_{n},
\end{eqnarray}

\begin{eqnarray}
\Theta_{nl}^{b} = A_{l}^{\ast} A_{n} + B_{l}^{\ast} B_{n} - C_{l}^{\ast} C_{n} - D_{l}^{\ast} D_{n},
\end{eqnarray}

\begin{eqnarray}
\Theta_{nl}^{c} = A_{l}^{\ast} A_{n} - B_{l}^{\ast} B_{n} + C_{l}^{\ast} C_{n} - D_{l}^{\ast} D_{n},
\end{eqnarray}

\begin{eqnarray}
\Theta_{nl}^{d} = A_{l}^{\ast} C_{n} + B_{l}^{\ast} D_{n} + C_{l}^{\ast} A_{n} + D_{l}^{\ast} B_{n},
\end{eqnarray}

\begin{eqnarray}
\Theta_{nl}^{e} = A_{l}^{\ast} B_{n} + B_{l}^{\ast} A_{n} + C_{l}^{\ast} D_{n} + D_{l}^{\ast} C_{n}.
\end{eqnarray}

By numerically solving these equations at each time $t$, one can calculate the values of $\dot{A}_{n}$, $\dot{B}_{n}$, $\dot{C}_{n}$, $\dot{D}_{n}$, $\dot{\mu}_{n}$, $\dot{\nu}_{n}$, and $\dot{\eta}_{nk}$ accurately. The fourth-order Runge-Kutta method is then adopted for the time evolution of the tunable Rabi dimer, including the time-dependent photon numbers, qubit polarization, and population of the bath modes.


\begin{thebibliography}{999}
\bibitem{Rabi1936}I. I. Rabi, Phys. Rev. \textbf{49}, 324--328 (1936).

\bibitem{Rabi1937}I. I. Rabi, Phys. Rev. \textbf{51}, 652--654 (1937).

\bibitem{braak_2016}D. Braak, Q.-H. Chen, M. T. Batchelor, and E. Solano, J. Phys. A Math. Theor. \textbf{49}, 300301 (2016).

\bibitem{alderete_2016}C. H. Alderete and B. M. Rodr\'iguez-Lara, J. Phys. A Math. Theor. \textbf{49}, 414001 (2016).

\bibitem{raimond_2001}J. M. Raimond, M. Brune, and S. Haroche, Rev. Mod. Phys. \textbf{73}, 565 (2001).

\bibitem{leibfried_2003}D. Leibfried, R. Blatt, C. Monroe, and D. Wineland, Rev. Mod.
Phys. \textbf{75}, 281 (2003).

\bibitem{scarlino_2015}P. Scarlino, E. Kawakami, D. R. Ward, D. E. Savage, M. G. Lagally, M. Friesen, S. N. Coppersmith, M. A. Eriksson, and L. M. K. Vandersypen, Phys. Rev. Lett. \textbf{115}, 106802 (2015).

\bibitem{chiorescu_2004}I. Chiorescu, P. Bertet, K. Semba, Y. Nakamura, C. J. P. M. Harmans, and J. E. Mooij, Nature \textbf{431}, 159 (2004).

\bibitem{zener_1932}C. Zener, Proc. R. Soc. London A \textbf{137}, 696 (1932).

\bibitem{landau_1932} L. D. Landau, Phys. Z. \textbf{2}, 46 (1932).

\bibitem{schevchenko_2012}S. N. Shevchenko, A. N. Omelyanchouk, and E. Il?ichev, Low Temp. Phys. \textbf{38}, 283 (2012).

\bibitem{temchenko_2011}E. A. Temchenko, S. N. Shevchenko, and A. N. Omelyanchouk, Phys. Rev. B 83, 144507 (2011).

\bibitem{fulu_2018}F. Zheng, Y. Zhang, L. Wang, Y. Wei, and Y. Zhao, Ann. Phys. \textbf{530}, 1800351 (2018).

\bibitem{saito_2006}K. Saito, M. Wubs, S. Kohler, P. H\"anggi, and Y. Kayanuma, Europhysics Lett. \textbf{76}, 22 (2006).

\bibitem{oliver_2005}W. D. Oliver, Y. Yu, J. C. Lee, K. K. Berggren, L. S. Levitov, and T. P. Orlando, Science (80-. ). 310, 1653 (2005).

\bibitem{niemcyzk_2010}T. Niemczyk, F. Deppe, H. Huebl, E. P. Menzel, F. Hocke, M. J. Schwarz, J. J. Garcia-Ripoll, D. Zueco, T. Hummer, E. Solano, A. Marx, and R. Gross, Nat Phys \textbf{6}, 772 (2010).

\bibitem{nalbach_2015}S. Javanbakht, P. Nalbach, and M. Thorwart, Phys. Rev. A \textbf{91}, 52103 (2015).

\bibitem{bishop_2010}L. S. Bishop, E. Ginossar, and S. M. Girvin, Phys. Rev. Lett. \textbf{105}, 100505 (2010)

\bibitem{henriet_2014}L. Henriet, Z. Ristivojevic, P. P. Orth, and K. Le Hur, Phys. Rev. A \textbf{90}, 23820 (2014).

\bibitem{wallraff_2004}A. Wallraff, D. I. Schuster, A. Blais, L. Frunzio, R.-S. Huang, J. Majer, S. Kumar, S. M. Girvin, and R. J. Schoelkopf, Nature \textbf{431}, 162 (2004).

\bibitem{Johansson_2009}J. Johansson, M. H. S. Amin, A. J. Berkley, P. Bunyk, V. Choi, R. Harris, M. W. Johnson, T. M. Lanting, S. Lloyd, and G. Rose, Phys. Rev. B \textbf{80}, 12507 (2009).

\bibitem{houck_2008}A. A. Houck, J. A. Schreier, B. R. Johnson, J. M. Chow, J. Koch, J. M. Gambetta, D. I. Schuster, L. Frunzio, M. H. Devoret, S. M. Girvin, and R. J. Schoelkopf, Phys. Rev. Lett. \textbf{101}, 80502 (2008).

\bibitem{viehmann_2013}O. Viehmann, J. von Delft, and F. Marquardt, Phys. Rev. Lett. \textbf{110}, 30601 (2013).

\bibitem{xiong_2015}H.-N. Xiong, P.-Y. Lo, W.-M. Zhang, D. H. Feng, and F. Nori, Sci. Rep. \textbf{5}, 13353 (2015).

\bibitem{egger_2013}D. J. Egger and F. K. Wilhelm, Phys. Rev. Lett. \textbf{111}, 163601 (2013).

\bibitem{zhou2015polaron} N. Zhou, Z. Huang, J. Zhu, V. Chernyak, and Y. Zhao, J. Chem. Phys. \textbf{143}, 014113 (2015).

\bibitem{huang_2017}Z. Huang, L. Chen, N. Zhou, and Y. Zhao, Ann. Phys. \textbf{529}, 1600367 (2017).

\bibitem{huang_lz_2018}Z. Huang and Y. Zhao, Phys. Rev. A \textbf{97}, 13803 (2018).

\bibitem{nalbach_2017}P. Nalbach, N. Klinkenberg, T.~Palm, and N. M\"uller, Phys. Rev. E \textbf{96}, 042134 (2017).

\bibitem{Braak2011}D.~Braak, Phys. Rev. Lett. \textbf{107}, 100401 (2011).

\bibitem{Zhong2017}H.~Zhong, Q.~Xie, X.~Guan, M. T. Batchelor, and C.~Lee, J. Phys. A Math. Theor. \textbf{50}, 113001 (2017).

\bibitem{WangLu2016}L.~Wang, L.~Chen, N.~Zhou, and Y.~Zhao, J. Chem. Phys., \textbf{144}, 024101 (2016).

\bibitem{zh_12}Y. Zhao, B. Luo, Y. Zhang, and J. Ye, J. Chem. Phys. \textbf{137}, 084113 (2012).

\bibitem{zh_97}Y. Zhao, D. W. Brown, and K. Lindenberg, J. Chem. Phys. \textbf{107}, 3159 (1997); \textbf{107}, 3179 (1997).

\bibitem{Zhou2016}N.~Zhou, L.~Chen, Z.~Huang, K.~Sun, Y.~Tanimura, and Y.~Zhao, J. Phys. Chem. A, \textbf{120}, 1562 (2016).

\bibitem{Chen2017}L.~Chen, R.~Borrelli, and Y.~Zhao, J. Phys. Chem. A, \textbf{121}, 8757-8770 (2017).

\bibitem{Wang2017}L.~Wang, Y.~Fujihashi, L.~Chen, and Y.~Zhao, J. Chem. Phys. \textbf{146}, 124127-124127 (2017).

\bibitem{huang_2017_off}Z. Huang, L. Wang, C. Wu, L. Chen, F. Grossmann, and Y. Zhao, Phys. Chem. Chem. Phys. \textbf{19}, 1655 (2017).

\bibitem{huang_SF_2017}Z. Huang, Y. Fujihashi, and Y. Zhao, J. Phys. Chem. Lett. \textbf{8}, 3306 (2017).

\bibitem{huang_2018_ac}Z. Huang, M. Hoshina, H. Ishihara, and Y. Zhao, Ann. Phys. \textbf{530}, 1800303 (2019).

\bibitem{Rossatto2016}D.\,Z. Rossatto, S.~Felicetti, H.~Eneriz, E.~Rico, M.~Sanz, and E.~Solano,  Phys. Rev. B, \textbf{93}(9), 094514 (2016).

\bibitem{WangYM2016}Y.~Wang, J.~Zhang, C.~Wu, J.\,Q. You, and G.~Romero, Phys. Rev. A, \textbf{94}(1), 012328 (2016).

\bibitem{Raftery2014}J.~Raftery, D.~Sadri, S.~Schmidt, E. T{\"u}reci, and A.\,A. Houck, Phys. Rev. X, \textbf{4}, 031043
  (2014).

\bibitem{Hwang2016}M.\,J. Hwang, M.\,S. Kim, and M.\,S. Choi, Phys. Rev. Lett. \textbf{116}, 153601 (2016).

\bibitem{hanggi_2005}K. M. Fonseca-Romero, S. Kohler, and P. H\"anggi, Phys. Rev. Lett. \textbf{95}, 140502 (2005).

\bibitem{savel_2012_2}S. E. Savel’ev, Z. Washington, A. M. Zagoskin, and M. J. Everitt, Phys. Rev. A \textbf{86}, 65803 (2012).

\bibitem{hau_2011}J. Hausinger and M. Grifoni, Phys. Rev. A \textbf{83}, 030301 (2011).

\bibitem{savel_2005}S. Savel'ev, A. L. Rakhmanov, and F. Nori, Phys. Rev. E \textbf{72}, 56136 (2005).

\bibitem{savel_2010}S. Savel'ev, A. M. Zagoskin, A. N. Omelyanchouk, and F. Nori, Chem. Phys. \textbf{375}, 180 (2010).

\bibitem{savel_2012_1}S. Savel'ev, A. M. Zagoskin, A. L. Rakhmanov, A. N. Omelyanchouk, Z. Washington, and F. Nori, Phys. Rev. A \textbf{85}, 013811 (2012).

\bibitem{son_2009}S.-K. Son, S. Han, and S.-I. Chu, Phys. Rev. A \textbf{79}, 32301 (2009).

\bibitem{cao_2007}X. Cao and H. Zheng, Phys. Rev. A \textbf{75}, 062121 (2007).

\bibitem{xia_2009}K. Xia, M. Macovei, J. Evers, and C. H. Keitel, Phys. Rev. B \textbf{79}, 24519 (2009).


\end{thebibliography}
\end{document}